\pdfoutput=1

\documentclass[aps,prb,reprint,showpacs,superscriptaddress,floatfix]{revtex4-1}

\usepackage{graphicx}
\usepackage{graphics}
\usepackage{amsmath}
\usepackage{amssymb}
\usepackage{amsfonts}
\usepackage{dcolumn}
\usepackage{dsfont}
\usepackage{latexsym}
\usepackage{rotating}
\usepackage{color}
\usepackage{latexsym}
\usepackage{bbm}
\usepackage{subfigure}
\usepackage{float}
\usepackage{epsfig}
\usepackage{epsf}
\usepackage{psfrag}
\usepackage{bm}
\usepackage{amsthm}
\usepackage{eucal}
\usepackage{mathrsfs}
\usepackage{url}
\usepackage{braket}
\usepackage{array}
\usepackage[utf8]{inputenc}
\usepackage{microtype}
\usepackage{multirow}

\usepackage{color} 

\usepackage{hyperref}
\hypersetup{
colorlinks=true,final=true,
        linkcolor=blue,
        citecolor=blue,
        filecolor=blue,
        urlcolor=blue,
}

\begin{document}

\title{Effects of Pressure on the Electronic and Magnetic Properties of Bulk NiI\texorpdfstring{$_{2}$}{2}}

\author{Jesse Kapeghian}
\email{jkapeghi@asu.edu}
\affiliation{Department of Physics, Arizona State University, Tempe, AZ - 85287, USA}

\author{Danila Amoroso}
\affiliation{Nanomat/Q-mat/CESAM, Universit\'{e} de Li\`{e}ge, B-4000 Sart Tilman, Belgium}

\author{Connor A. Occhialini}
\affiliation{Department of Physics, Massachusetts Institute of Technology, Cambridge, MA 02139, USA}

\author{Luiz G. P. Martins}
\affiliation{Department of Physics, Massachusetts Institute of Technology, Cambridge, MA 02139, USA}

\author{Qian Song}
\affiliation{Department of Physics, Massachusetts Institute of Technology, Cambridge, MA 02139, USA}

\author{Jesse S. Smith}
\affiliation{HPCAT, Advanced Photon Source, Argonne National Laboratory, Lemont, IL 60439, USA}

\author{Joshua J. Sanchez}
\affiliation{Department of Physics, Massachusetts Institute of Technology, Cambridge, MA 02139, USA}

\author{Jing Kong}
\affiliation{Department of Electrical Engineering and Computer Science, Massachusetts Institute of Technology, Cambridge, MA 02139, USA}

\author{Riccardo Comin}
\affiliation{Department of Physics, Massachusetts Institute of Technology, Cambridge, MA 02139, USA}

\author{Paolo Barone}
\affiliation{Consiglio Nazionale delle Ricerche CNR-SPIN, Area della Ricerca di Tor Vergata, Via del Fosso del Cavaliere 100, I-00133 Rome, Italy}

\author{Bertrand Dup\'{e}}
\affiliation{Nanomat/Q-mat/CESAM, Universit\'{e} de Li\`{e}ge, B-4000 Sart Tilman, Belgium}
\affiliation{Fonds de la Recherche Scientifique (FNRS), Bruxelles, Belgium}

\author{Matthieu J. Verstraete}
\affiliation{Nanomat/Q-mat/CESAM, Universit\'{e} de Li\`{e}ge, B-4000 Sart Tilman, Belgium}
\affiliation{European Theoretical Spectroscopy Facility www.etsf.eu}

\author{Antia S. Botana}
\email{antia.botana@asu.edu}
\affiliation{Department of Physics, Arizona State University, Tempe, AZ - 85287, USA}

\date{\today}

\begin{abstract}

Transition metal dihalides have recently garnered interest in the context of two-dimensional van der Waals magnets as their underlying geometrically frustrated triangular lattice leads to interesting competing exchange interactions. In particular, NiI$_{2}$ is a  magnetic semiconductor that has been long known for its exotic helimagnetism in the bulk. Recent experiments have shown that the helimagnetic state survives down to the monolayer limit with a layer-dependent magnetic transition temperature that suggests a relevant role of the interlayer coupling. Here, we explore the effects of hydrostatic pressure as a means to  enhance this interlayer exchange and ultimately tune the electronic and magnetic response of NiI$_{2}$. 
We study first the evolution of the structural parameters as a function of external pressure using first-principles calculations combined with x-ray diffraction measurements. We then examine the evolution of the electronic structure and magnetic exchange interactions  via first-principles calculations and Monte Carlo simulations. We find that the leading interlayer coupling is an antiferromagnetic second-nearest neighbor interaction that increases monotonically with pressure. The ratio between isotropic third- and first-nearest neighbor intralayer exchanges, which controls the magnetic frustration and determines the magnetic propagation vector $\mathbf{q}$ of the helimagnetic ground state, is also enhanced by pressure. As a consequence, our Monte Carlo simulations show a monotonic increase in the magnetic transition temperature, indicating that pressure is an effective means to tune the magnetic response of NiI$_{2}$.

\end{abstract}

\maketitle

\section{Introduction} \label{intro}


Magnetic two dimensional (2D) van der Waals (vdW) materials have attracted much attention \cite{Blei_APR_2021} since Ising-type magnetic orders were demonstrated in the monolayer limit for antiferromagnetic FePS$_3$\cite{Wang_2DM_2016,Lee_NL_2016} and for ferromagnetic CrI$_3$ \cite{Huang_Nat_2017}. This series of discoveries naturally led researchers to explore the possibility of realizing both electric and magnetic orders simultaneously in a vdW material down to the single-layer limit. In this context, transition metal (TM) dihalides are emerging as novel platforms to explore magnetoelectricity and noncollinear spin textures in 2D. Indeed, recent experiments \cite{Song_Nat_2022} in NiI$_{2}$ have revealed that multiferroicity (MF) in this compound persists from the bulk to the monolayer limit leading to the realization of a vdW material that simultaneously displays magnetic and ferroelectric order.

\begin{figure}
\includegraphics[width=\columnwidth]{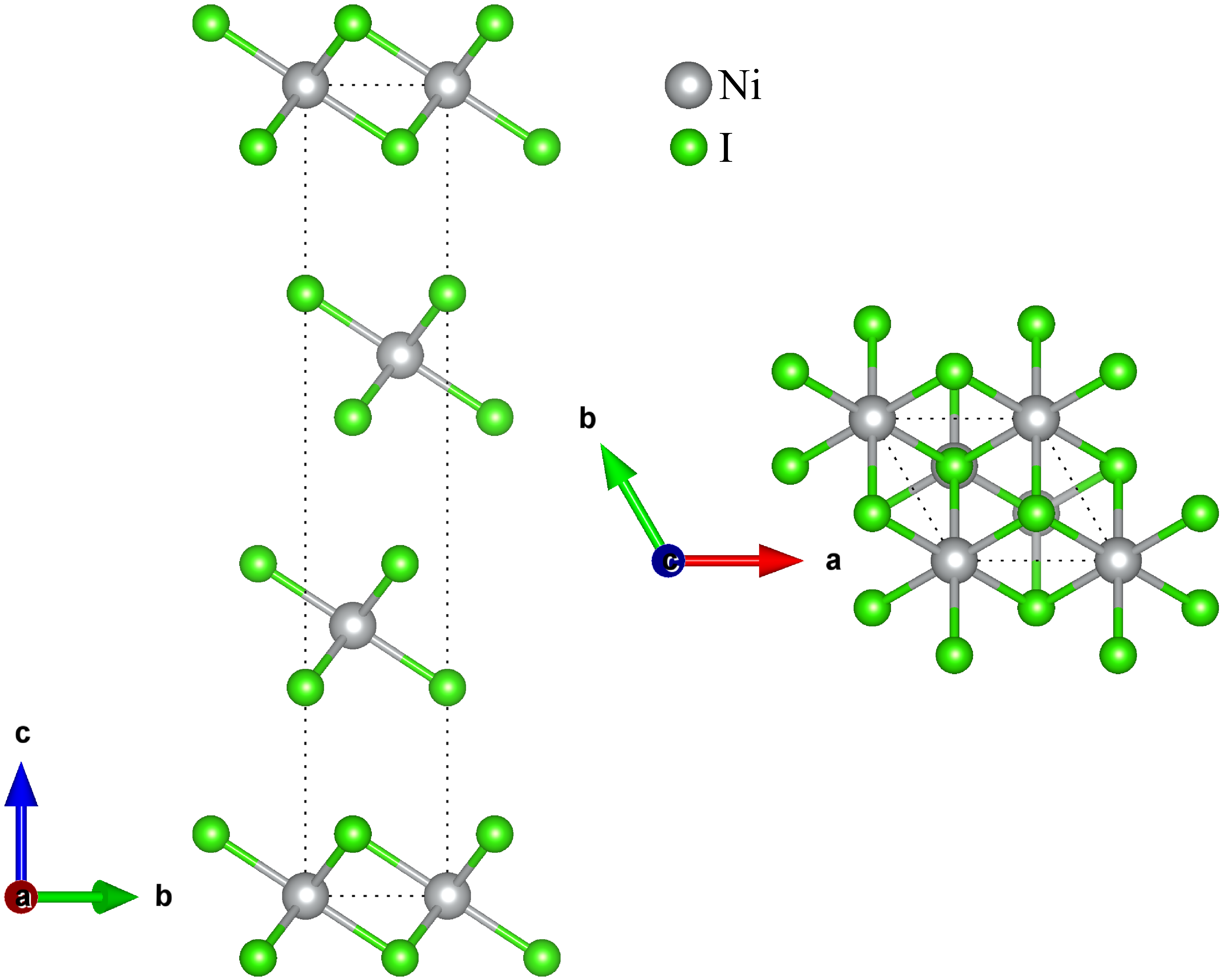}
\caption{Crystal structure of bulk NiI$_{2}$ in the rhombohedral ($R \overline{3} m$) phase showing the out-of-plane (left) and in-plane (right) arrangement of atoms, where gray (green) spheres represent Ni (I) atoms and the unit cell boundary is marked by a black dotted line.}
\label{fig:1}
\end{figure}

Structurally, bulk NiI$_{2}$ adopts the rhombohedral ($R \overline{3} m$) CdCl$_{2}$ crystal lattice at room temperature, containing triangular nets of TM (Ni) cations in edge sharing octahedral coordination and forming NiI$_{2}$ layers separated by vdW gaps  \cite{Nasser_SSC_1992,Ketelaar_ZfK_1934,McGuire_Cryst_2017} (see Fig. \ref{fig:1}). 
Bulk NiI$_{2}$ 
is known to undergo two phase transitions upon cooling; the first is to an antiferromagnetic (AFM) state with a N\'{e}el temperature $T_{\mathrm{N},1} \simeq 75$ K and the second transition is to a single-\textbf{q} proper-screw helimagnetic (HM) ground state at $T_{\mathrm{N},2} \simeq 60$ K \cite{Adam_SSC_1980,Day_JPC_1976,Day_JPC_1980,Billerey_PLA_1977,Kuindersma_PBC_1981,McGuire_Cryst_2017}. 
The $T_{\mathrm{N},2}$ transition is concomitant with a lowering of the crystal symmetry from rhombohedral to monoclinic \cite{Kuindersma_PBC_1981,Liu_ACSN_2020}. In addition, $T_{\mathrm{N},2}$ also marks the onset of type-II MF order \cite{Kurumaji_PRB_2013,Song_Nat_2022} as it was found that the non-collinear magnetic state hosts a ferroelectric polarization tunable via a magnetic field. As mentioned above, the HM state in NiI$_{2}$ is found to be robust down to the single-layer limit with experiments indicating that $T_{\mathrm{N},2}$ decreases monotonically with the number of layers \cite{Song_Nat_2022}. This continuous decrease of $T_{\mathrm{N},2}$  indicates that the interlayer exchange coupling plays an important role in the magnetic order of NiI$_{2}$ \cite{Song_Nat_2022}.

Exploring ways to enhance the interlayer exchange in NiI$_{2}$, in order to tune its electronic and magnetic response, is then a natural path to pursue. Here, we study the effects of hydrostatic pressure as a means to achieve this goal, using first-principles calculations in conjunction with high-pressure x-ray diffraction (XRD) experiments. We present a systematic density-functional theory (DFT)-based study of the magnetic interactions of NiI$_{2}$ under pressures up to 20 GPa. 
We find that pressure has a significant effect on the interlayer coupling, but also in some of the leading intralayer interactions. Specifically, while the dominant intralayer exchange at ambient pressure (ferromagnetic nearest-neighbor $J^{\parallel 1}$) is weakly pressure dependent, there is a significant enhancement of the antiferromagnetic third neareast-neighbor intralayer interaction ($J^{\parallel 3}$) and of the antiferromagnetic interlayer exchange ($J^{\perp \mathrm{eff}}$). Monte Carlo (MC) simulations reveal a 3-fold increase in the helimagnetic transition temperature between 0 and 10 GPa followed by saturation at higher pressures. 

\section{Computational Methods} \label{comp}

\subsection{First-Principles Calculations} \label{DFT}

We performed DFT-based calculations using the projector augmented wave (PAW) method \cite{Kresse_PRB_1999} as implemented in the VASP code \cite{Kresse_PRB_1996,Kresse_CMS_1996}. The wave functions were expanded in the plane-wave basis with a kinetic-energy cut-off of 500 eV. Consistent with our previous work (see Refs. [\onlinecite{Amoroso_NC_2020, Song_Nat_2022}]), the 3$p$, 3$d$, and 4$s$ orbitals (3$p$$^{6}$3$d$$^{9}$4$s$$^{1}$ configuration) were considered as valence states for the Ni atoms while for the I atoms the 5s and 5p orbitals (5$s^{2}$5$p^{5}$ configuration) were as considered valence states.

Hydrostatic pressure was applied in increments of 5 GPa during relaxation for pressures up to 20 GPa. The structural degrees of freedom considered during the optimization of the bulk unit cells at each pressure were atomic positions, cell shape, and cell volume, and we restricted our analysis to the rhombohedral phase. Tolerances for energy and force minimization during relaxation were set at $10^{-10}$ eV and $10^{-3}$ eV/\AA, respectively. 
Different DFT functionals and spin configurations were tested, after which the structural parameters were compared to those obtained from experimental XRD data. The best agreement was obtained for an AFM state (consisting ferromagnetic (FM) planes coupled AFM out-of-plane) using the Perdew-Burke-Ernzerhof (PBE) \cite{Perdew_PRL_1996} version of the generalized gradient approximation (GGA) functional with the DFT-D3 van der Waals correction \cite{Grimme_JCP_2010}, and including an on-site Coulomb repulsion $U$ using the Liechtenstein \cite{Liechtenstein_PRB_1995} approach in order to account for correlation effects in the Ni-$d$ electrons \cite{Rohrbach_JPCM_2003}. The $U$ and Hund's coupling $J_{\mathrm{H}}$ values used (3.24 and 0.68 eV, respectively) were taken from constrained random phase approximation (cRPA) calculations \cite{Riedl_PRB_2022}. To accommodate the AFM order, a $1 \times 1 \times 2$ supercell was used and the sampling over the Brillouin zone (BZ) was performed with a $36 \times 36 \times 3$ Monkhorst-Pack $k$-mesh centered on $\Gamma$.  Note that this AFM order is consistent with the $c$-component of the magnetic propagation vector, that is $\sim$ 3/2 not only for NiI$_{2}$ but across the Ni-dihalide series \cite{McGuire_Cryst_2017}. Electronic structure calculations for the optimized bulk structures at each pressure were performed with a  tolerance of $10^{-8}$ eV for the electronic energy minimization. 

Consistently with Ref. [\onlinecite{Amoroso_NC_2020}], we employed the four-state method (explained in detail in Refs. [\onlinecite{Xiang_DT_2013,Xiang_PRB_2011,Sabani_PRB_2020,Xu_NPJCM_2018,Xu_PRB_2020}]) to calculate the exchange couplings and anisotropies for NiI$_{2}$. The
four-state method is based on total energy mapping through noncollinear, magnetic DFT calculations, including spin-orbit coupling (SOC).
Each magnetic interaction parameter is related to the energies
of four distinct magnetic configurations
wherein the directions of the magnetic moments were constrained and large supercells were used to inhibit coupling between distant neighbors. At each pressure, intralayer (interlayer) magnetic constants were calculated using monolayer (bilayer) structures built from the relaxed bulk structures. In both the mono- and bilayers, we used a distance of more than 20 \AA \, with respect to the periodic repetition along the out-of-plane direction. For the monolayers, a $5 \times 4 \times 1$ supercell was used for first- and second-nearest in-plane neighbors (and single-ion anisotropy) and a $6 \times 3 \times 1$ supercell was used for third-nearest in-plane neighbors. For the bilayer, a $3 \times 3 \times 1$ supercell was used for first-, second-, and third-nearest out-of-plane neighbors. In all mono- and bilayer cases, a $\Gamma$-centered $6 \times 6 \times 2$ $k$-mesh was employed for BZ sampling.

\begin{figure*}
\includegraphics[width=2\columnwidth]{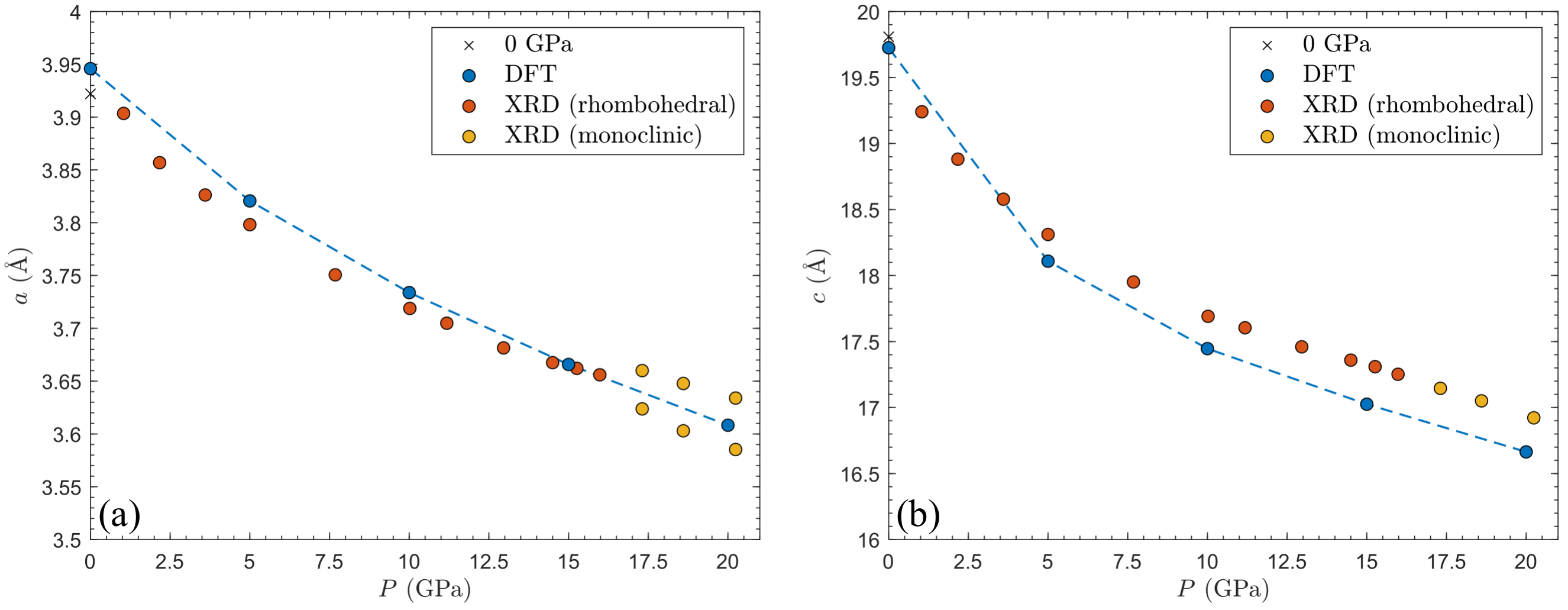}
\caption{In-plane (a) and out-of-plane (b) relaxed lattice parameters for bulk NiI$_{2}$ from DFT (blue circles) along with our experimental XRD data (orange/yellow circles for rhombohedral/monoclinic phases) taken at various pressures at 200 K. The black crosses represent ambient pressure experimental values obtained from neutron diffraction experiments at 300 K: $a=3.922$ \AA, $c=19.808$ \AA \, \cite{Kuindersma_PBC_1981}. 
}
\label{fig:2}
\end{figure*}

\subsection{Monte Carlo Simulations} \label{MC}

We used the simulation code Matjes\cite{matjes} to perform Monte Carlo calculations and extract the critical temperatures.
Specifically, up to $\sim 10^{6}$ thermalization steps were used at each simulated temperature followed by $10^{4}$ MC steps for statistical averaging. At each temperature, the average total energy, magnetization, and specific heat were calculated. Further, a standard Metropolis algorithm was used on 
supercells of size $L \times L \times 4$ with periodic boundary conditions. For each bulk structure optimized at a different pressure, the supercell size $L$ was chosen according to $L \simeq nL_{\mathrm{m.u.c.}}$ where $n$ is an integer and $L_{\mathrm{m.u.c.}}$ is the minimum lateral size of the magnetic unit cell which is required to faithfully represent the noncollinear spin configuration of the HM ground state. The magnetic unit cell length can be estimated as $L_{\mathrm{m.u.c.}} \sim 4 \pi /q$, where $q$ is the magnitude of the magnetic propagation vector $\mathbf{q}$ which minimizes the exchange interaction energy in momentum space. Considering an isotropic model where the second nearest-neighbor interaction is neglected, an analytical solution can be obtained for the wave vector, $q= 2\arccos{\left[ \left( 1 + \sqrt{1 - 2 (J^{\parallel 1}/J^{\parallel 3}) } \right) / 4 \right]}$ \cite{Hayami_PRB_2016,Batista_RPP_2016}. 

\section{Results} \label{results}

\subsection{Structural Optimizations and Electronic Structure} \label{struct}

Fig. \ref{fig:2} shows the optimized bulk lattice parameters calculated at pressures up to 20 GPa via first-principles calculations. These lattice parameters were derived within an AFM state (consisting of FM planes coupled AFM out-of-plane) using the GGA-PBE functional including a DFT-D3 correction and an on-site Coulomb repulsion $U = 3.24$ eV. 
Computational results are compared with experimental values in the same pressure range. The ambient pressure experimental structural data are taken from Ref. [\onlinecite{Kuindersma_PBC_1981}] (at 300 K) while for finite pressures the data are obtained from our XRD experiments performed at 200 K (see Appendix \ref{AppA} for further details). Up to $\sim$15 GPa the experimental in-plane lattice parameters (orange circles) are equivalent indicating rhombohedral symmetry (i.e. $a=b$), while above $\sim$15 GPa there is a splitting in the in-plane lattice parameter (yellow circles) indicating a symmetry-lowering to a monoclinic phase ($a \neq b$). 
DFT calculations are restricted to the rhombohedral phase, as explained above. The DFT-derived optimized lattice parameters show in-plane lattice constants ($a=b$)
that exhibit a monotonic decrease with increasing pressure (from $\sim 3.95$ \AA \, at ambient pressure to $\sim 3.61$ \AA \, at 20 GPa), in good agreement with the experimental data (Fig. \ref{fig:2}a). 
An expected monotonic decrease is also observed in the out-of-plane optimized lattice parameter ($c$) with increasing pressure (from $\sim 19.72$ \AA \, at ambient pressure to $\sim 16.66$ \AA \, at 20 GPa) with the experimental data showing the same trends (Fig. \ref{fig:2}b). As expected for a vdW material, the structural response of NiI$_{2}$ to pressure is highly anisotropic with the $c$ lattice parameter showing a rate of compression that is roughly double that of the $a$ lattice parameter. For additional optimization calculations using various computational methods, see Appendix \ref{AppA}.


\begin{figure*}
\includegraphics[width=2\columnwidth]{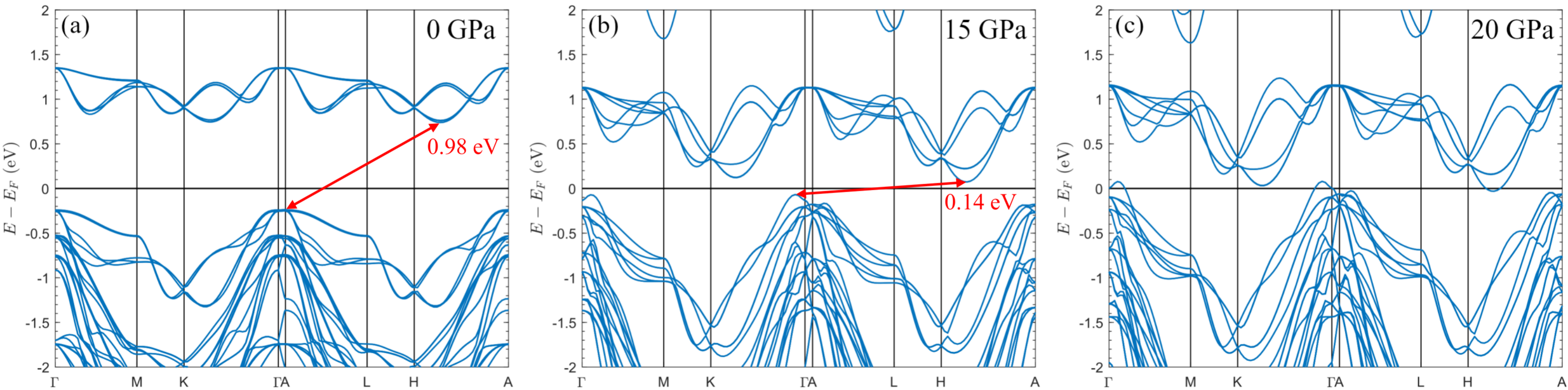}
\caption{GGA-PBE+D3+$U$ (3.24 eV) band structure plots for bulk NiI$_{2}$ (calculated with AFM order) at ambient pressure (a), 15 GPa (b) and 20 GPa (c) where the energy band gaps are indicated in both cases. Reciprocal space coordinates: $\Gamma=(0,0,0)$, $\mathrm{M}=(1/2,0,0)$, $\mathrm{K}=(1/3,1/3,0)$, $\mathrm{A}=(0,0,1/2)$, $\mathrm{L}=(1/2,0,1/2)$, $\mathrm{H}=(1/3,1/3,1/2)$ (see Appendix \ref{AppB} for the schematic of the BZ showing these high-symmetry points).}
\label{fig:3}
\end{figure*}


Fig. \ref{fig:3} shows the evolution of the corresponding band structure of NiI$_{2}$ upon pressure for the optimized structures described above (at 0, 15, and 20 GPa). The insulator-to-metal transition can be clearly observed: an insulating solution is obtained up to 15 GPa and a metallic state at 20 GPa, consistent with the experimentally reported insulator-to-metal transition at 19 GPa \cite{Pasternak_AIPCP_1994} (see Appendix \ref{AppB} for a full evolution of the band structure from 0-20 GPa). 
Note that 
along the $\Gamma$ to A directions some bands are quite dispersive indicating a relatively large degree of interaction between layers.  
This out-of-plane dispersion increases with pressure: that is, as the interlayer distance decreases. Orbital-resolved densities of states (DOS) along with band character plots indicate that the dispersion is driven primarily from the I-$p_{z}$ states (see Appendix \ref{AppB}). 


\subsection{Exchange Interactions} \label{xc}

The relevant magnetic interactions between localized spins ($\mathbf{S}_{i}$) in NiI$_{2}$ are the intra- and interlayer exchange couplings and the single-ion anisotropy. The microscopic model which encapsulates the in-plane interactions is given by the 2D anisotropic Heisenberg Hamiltonian used in Ref. [\onlinecite{Amoroso_NC_2020}] to describe monolayer NiI$_{2}$, to which we have added an isotropic interlayer exchange term to account for the magnetic interactions between layers (we only consider the isotropic component for the out-of-plane exchange since the short-range anisotropic contributions arising from the spin-orbit interaction are negligible in comparison). 
Therefore, the corresponding Hamiltonian for our problem can be broken up into in-plane and out-of-plane components: $H = H^{\parallel} + H^{\perp}$, where
\begin{equation} \label{eq:1}
    H^{\parallel} = \frac{1}{2} \sum_{i \neq j} \mathbf{S}_{i} \cdot \mathbf{J}_{ij}^{\parallel} \cdot \mathbf{S}_{j} + \sum_{{i}} \mathbf{S}_{i} \cdot \mathbf{A}_{i} \cdot \mathbf{S}_{i}
\end{equation}
represents the in-plane exchange (including isotropic and anisotropic coupling) and single-ion terms while
\begin{equation} \label{eq:2}
    H^{\perp} = \frac{1}{2} \sum_{i,j} J_{ij}^{\perp} \, \mathbf{S}_{i} \cdot \mathbf{S}_{j}
\end{equation}
contains the exchange (isotropic only) between layers \cite{Amoroso_NC_2020}. Here, $\mathbf{J}_{ij}^{\parallel}$ is the intralayer exchange interaction tensor, $J_{ij}^{\perp}$ is the isotropic interlayer exchange constant, and $\mathbf{A}_{i}$ is the single-ion anisotropy (SIA) tensor. 
The intralayer exchange interaction tensor is made up of contributions from isotropic and symmetric anisotropic (also called two-site anisotropy, or TSA) components. The antisymmetric exchange (Dzyaloshinskii–Moriya (DM)) interaction vanishes due to the presence of inversion symmetry \cite{Amoroso_NC_2020,Moriya_PR_1960,Simon_JPCM_2014}. 
The indices $i,j$ denote Ni atom sites wherein we consider up to third nearest neighbor isotropic exchanges both in-plane (Eq. \ref{eq:1}) and out-of-plane (Eq. \ref{eq:2}) (see Appendix \ref{AppC} for schematic representations of these leading interactions). The full tensor has been taken into account for the in-plane nearest neighbor exchange interaction. 
The single-ion anisotropy in Eq. \ref{eq:1} is an on-site term.
The factors of 1/2 in front of the exchange terms account for double-counting. Note the sign conventions in Eqs. \ref{eq:1} and \ref{eq:2} implying that a positive (negative) isotropic exchange interaction favors an antiparallel (parallel) alignment of spins and a positive (negative) scalar single-ion parameter indicates an easy-plane (easy-axis) anisotropy. Further details of the derivation of the magnetic interactions are given in Appendix \ref{AppC}.

\renewcommand{\arraystretch}{1.3}
\begin{table}
\centering
\begin{tabular}{m{3.75em} m{3.75em} m{3.75em} m{3.75em} c}
\hline
\hline
\multicolumn{5}{c}{Isotropic intralayer exchanges} \\
$P$ & $J^{\parallel 1}$ & $J^{\parallel 2}$ & $J^{\parallel 3}$ & $J^{\parallel 3}/J^{\parallel 1}$ \\
\hline
0 & $-$4.54 & $-$0.19 & +3.68 & $-$0.81 \\
5 & $-$4.97 & $-$0.30 & +5.85 & $-$1.18 \\
10 & $-$5.04 & $-$0.51 & +8.11 & $-$1.61 \\
15 & $-$4.82 & $-$0.92 & +10.43 & $-$2.16 \\
\end{tabular}
\begin{tabular}{lcccccc}
\hline
\multicolumn{7}{c}{SIA and intralayer TSA} \\
$P$ & $A$ & $J^{\mathrm{S} \parallel 1}_{xx}$ & $J^{\mathrm{S} \parallel 1}_{yy}$ & $J^{\mathrm{S} \parallel 1}_{zz}$ & $J^{\mathrm{S} \parallel 1}_{yz}$ & $J^{\mathrm{S} \parallel 1}_{yz}/J^{\parallel 1}$ \\
\hline
0 & +0.22 & $-$0.62 & +0.74 & $-$0.13 & $-$0.85 & +0.19 \\
5 & +0.26 & $-$0.66 & +0.82 & $-$0.16 & $-$0.96 & +0.19 \\
10 & +0.25 & $-$0.68 & +0.89 & $-$0.21 & $-$1.03 & +0.20 \\
15 & +0.24 & $-$0.65 & +0.95 & $-$0.29 & $-$1.07 & +0.22 \\
\hline
\end{tabular}
\begin{tabular}{m{2.78em} m{2.78em} m{2.78em} m{2.78em} m{2.78em} c}
\multicolumn{6}{c}{Isotropic interlayer exchanges} \\
$P$ & $J^{\perp 1}$ & $J^{\perp 2}$ & $J^{\perp 3}$ & $J^{\perp \mathrm{eff}}$ & $J^{\perp \mathrm{eff}}/J^{\parallel 1}$ \\
\hline
0 & $-$0.08 & +1.48 & +0.52 & +2.44 & $-$0.54 \\
5 & +0.06 & +4.83 & +1.47 & +7.82 & $-$1.57 \\
10 & +0.35 & +7.80 & +1.78 & +11.70 & $-$2.32 \\
15 & +0.59 & +10.55 & +1.15 & +13.44 & $-$2.79 \\
\hline
\hline
\end{tabular}
\caption{Bulk NiI$_{2}$ intra- (top panel) and interlayer (bottom panel) isotropic exchange interactions, plus SIA ($A$), and first-nearest neighbor  in-plane TSA constants  (middle panel) in the cartesian ${x, y, z}$ reference where $x$ was chosen to be parallel to the Ni-Ni bonding vector for different pressures ($P$) calculated via the four-state method. $J^{\perp \mathrm{eff}}$ represents the effective interlayer exchange and is obtained as $J^{\perp 1} + J^{\perp 2} + 2J^{\perp 3}$, where the coefficient in the last term arises because there are twice as many out-of-plane third nearest-neighbors as first- and second nearest-neighbors. Pressure is in units of GPa and exchange values are given in units of meV.}
\label{table:1}
\end{table}

\begin{figure}
\includegraphics[width=\columnwidth]{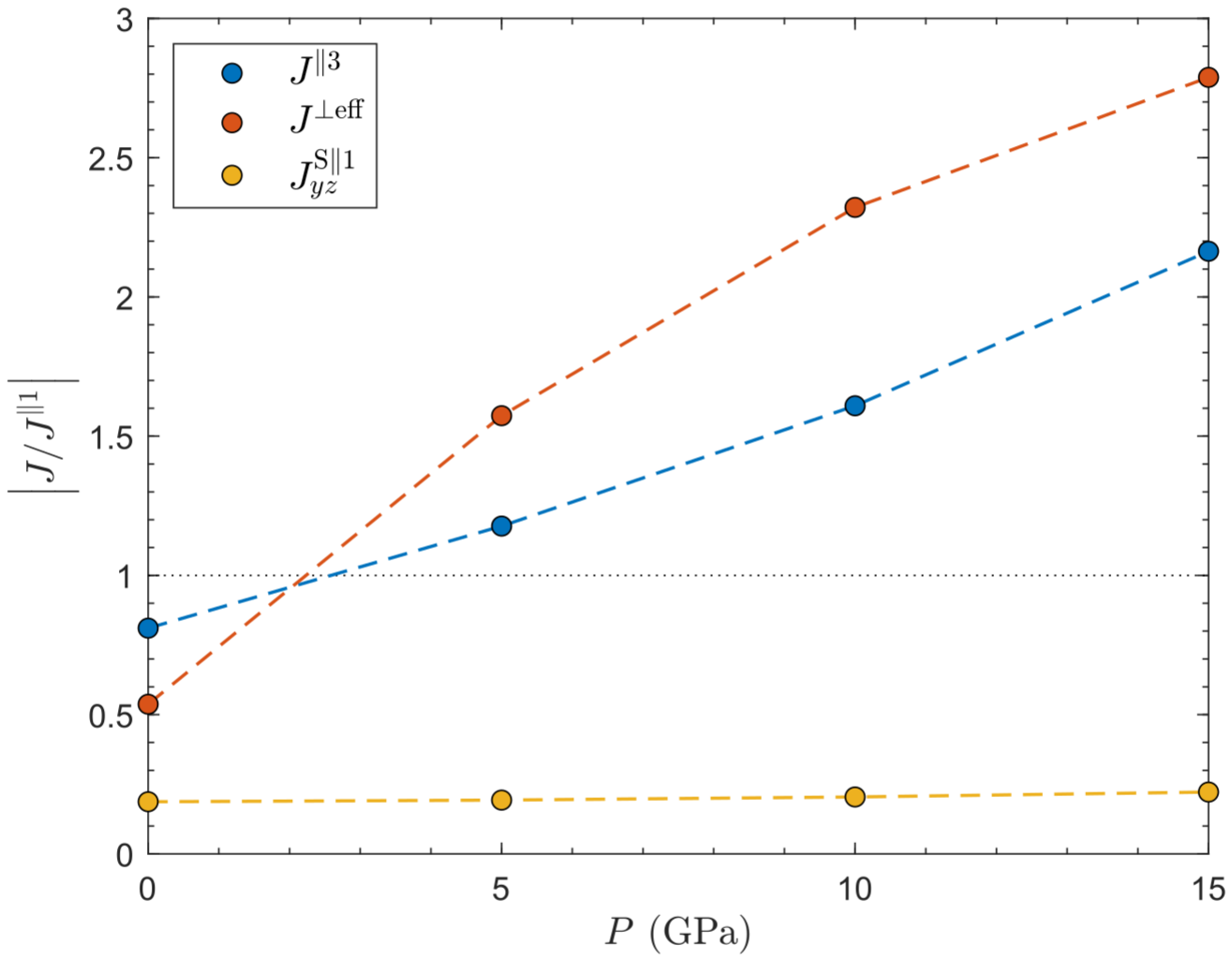}
\caption{Ratio of exchange couplings to $J^{\parallel 1}$ at different pressures in bulk NiI$_{2}$.}
\label{fig:4}
\end{figure}

Table \ref{table:1} contains the derived intralayer and interlayer coupling constants for NiI$_{2}$ 
calculated via the four-state method for pressures up to 15 GPa (20 GPa results were not included since at this pressure we obtain a metallic solution). As described before, at ambient pressure the non-collinear magnetic ground state of NiI$_{2}$ is realized via the competition between the dominant intralayer FM first-nearest neighbor exchange ($J^{\parallel 1}$) and the AFM third-nearest neighbor exchange ($J^{\parallel 3}$). This competition (measured by the ratio $J^{\parallel 3}/J^{\parallel 1} = -0.81$) results in a strong magnetic frustration  which favors helimagnetic phases \cite{Amoroso_NC_2020, Rastelli_PBC_1979}. One other important quantity is the ratio $J^{\mathrm{S} \parallel 1}_{yz}/J^{\parallel 1} = 0.19$, that measures the canting of the two-site anisotropy axes from the direction perpendicular to the layers. Finally, the out-of-plane modulation of the magnetic propagation vector in the bulk is determined by a net AFM interlayer exchange $J^{\perp \mathrm{eff}}$ \cite{Song_Nat_2022} which we now decompose further revealing the dominance of the second nearest-neighbor term ($J^{\perp 2}$). 

With pressure, the signs of the dominant intralayer isotropic exchange interactions remain: $J^{\parallel 1}$ is FM while $J^{\parallel 3}$ is AFM for all applied pressure values. However, while $J^{\parallel 1}$ is weakly pressure dependent, $J^{\parallel 3}$ significantly increases with pressure, becoming the dominant exchange already at 5 GPa. 
$J^{\parallel 2}$ remains small in comparison to both $J^{\parallel 1}$ and $J^{\parallel 3}$. As shown in Ref. \onlinecite{Riedl_PRB_2022}, $J^{\parallel 1}$ and $J^{\parallel 2}$ both comprise two main contributions, one being FM (mostly mediated by I-$p$ states) and the other one AFM (arising mostly from a direct $d-d$ overlap between $t_{2g}$-like states). Their partial compensation is likely linked to the weak sensitivity of $J^{\parallel 1}$ to pressure. Instead,  $J^{\parallel 3}$ only portrays AFM contributions arising from $e_{g}$-$p$ hybridizations \cite{Riedl_PRB_2022}, explaining its monotonic dependence with pressure. In this manner, the $J^{\parallel 3}/J^{\parallel 1}$ ratio changes from -0.81 at ambient pressure to -2.16 at 15 GPa (see Fig. \ref{fig:4}). Note that the ratio $\big| J^{\parallel 3}/J^{\parallel 1} \big|$ is essential for determining the incommensurate helimagnetic propagation vector \cite{Kuindersma_PBC_1981,Rastelli_PBC_1979}. In particular, the observed increase in $\big| J^{\parallel 3}/J^{\parallel 1} \big|$ with pressure would favor a shorter in-plane spiral pitch, i.e. a larger in-plane $\mathbf{q}_{\parallel}$, with larger nearest-neighbor spin angle (see Appendix \ref{AppC}).
The single-ion (easy-plane) anisotropy and the intralayer anisotropic exchanges do not significantly change with pressure, with the $J^{\mathrm{S} \parallel 1}_{yz}/J^{\parallel 1}$ ratio remaining almost constant (and small) up to 15 GPa, as shown in Fig. \ref{fig:4}. 

\begin{figure*}
\includegraphics[width=2\columnwidth]{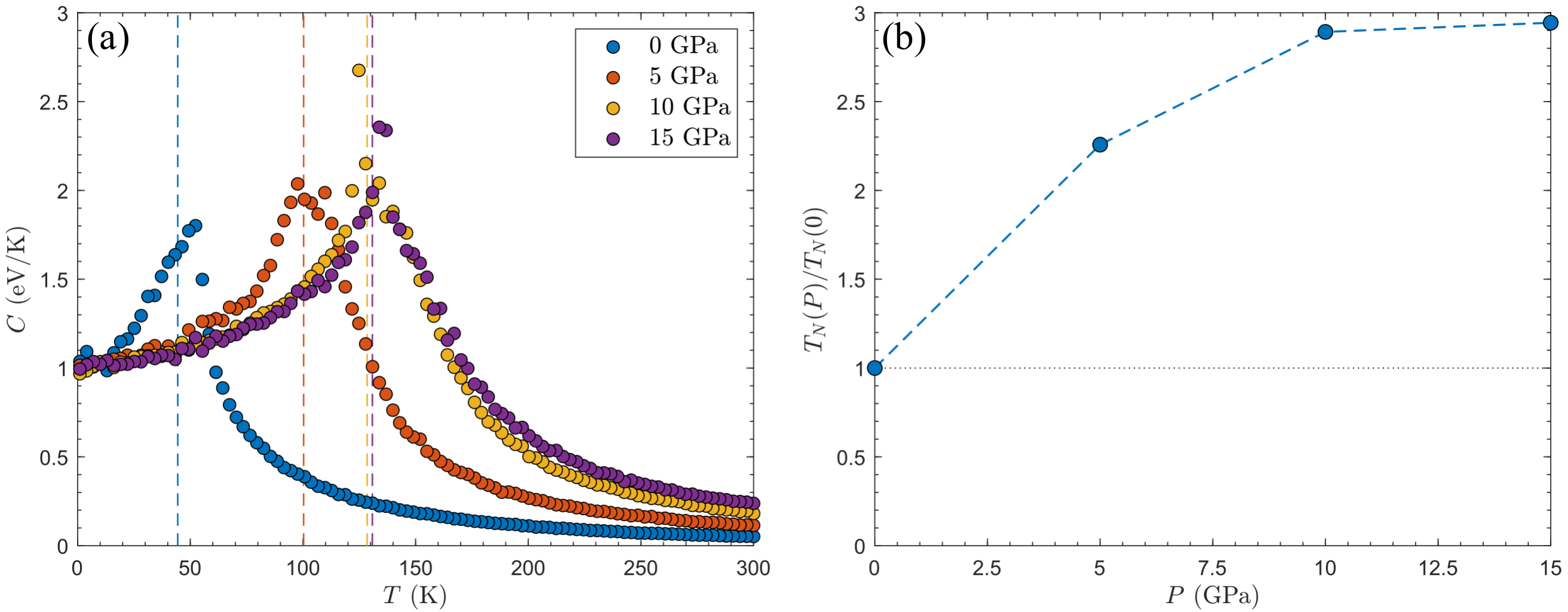}
\caption{(a) Bulk NiI$_{2}$ specific heat $C$ as a function of temperature $T$ for pressures of 0, 5, 10, and 15 GPa (blue, orange, yellow, and purple circles, respectively) obtained from Monte Carlo simulations. The dashed lines indicate the N\'{e}el temperature at each pressure (obtained via curve fitting with a general Lorentzian function): 44.4, 100.3, 128.5, and 130.8 K for 0, 5, 10, and 15 GPa, respectively. (b) Ambient-pressure-normalized N\'{e}el temperature values for bulk NiI$_{2}$ from MC calculations (blue circles) as a function of pressure $P$.}
\label{fig:5}
\end{figure*}

Concerning the interlayer exchanges, they increase significantly as a consequence of the large decrease of the $c$ lattice parameter with pressure, as expected in a vdW material. AFM $J^{\perp 2}$ remains the dominant interlayer interaction: 
at 10 GPa it actually becomes the second-largest interaction overall (behind $J^{\parallel 3}$) and at 15 GPa it even slightly surpasses $J^{\parallel 3}$ to become the dominant exchange interaction. 
The effective interlayer exchange $J^{\perp \mathrm{eff}}$ 
also increases monotonically with pressure, as does the magnitude of the ratio $J^{\perp \mathrm{eff}}/J^{\parallel 1}$ (see Fig. \ref{fig:4}). The role of the interlayer interaction in stabilizing the helimagnetic transition under pressure can be regarded as a complementary effect to its role in enhancing the transition when going from the monolayer to the bulk \cite{Song_Nat_2022}.



\subsection{Transition Temperatures} \label{TN}

The magnetic constants derived from the four-state method (Table \ref{table:1}) along with the optimized structural parameters (Fig. \ref{fig:2}) were used in Monte Carlo simulations to obtain the specific heat over a temperature range of 1 to 300 K. For each pressure, the ($T_{\mathrm{N},2}$) temperature was then extracted from the specific heat plots via curve fitting to a general Lorentzian function (see Appendix \ref{AppD}). Fig. \ref{fig:5}a shows the bulk NiI$_{2}$ specific heat vs. temperature plots from ambient pressure up to 15 GPa, where $T_{\mathrm{N},2}$ for each pressure is indicated by a dashed vertical line. 
$T_{\mathrm{N},2}$ is found to exhibit an almost 3-fold increase between 0 and 10 GPa (from 44 to 128 K), followed by saturation at higher pressure. Fig. \ref{fig:5}b shows the  $T_{\mathrm{N},2}$ from our MC calculations as a function of pressure, where the data points are normalized with respect to the calculated ambient pressure value ($T_{\mathrm{N},2}(0) = $ 44.4 K). 
The saturation of $T_{\mathrm{N},2}$ above 10 GPa is indicative of the competition between intra- and interlayer interactions (see Appendix \ref{AppE}). 


MC simulations were also obtained (using the same magnetic parameters and optimized structural data) for mono- and bilayer NiI$_{2}$ where it was found that the $T_{\mathrm{N}}$ values as a function of pressure behaved linearly beyond 5 GPa, further supporting the claim that a larger interlayer coupling acts to reduce $T_{\mathrm{N}}$ (see Appendix \ref{AppD}). Furthermore, the expected increase in $T_{\mathrm{N}}$ with number of layers was observed.

Overall, our MC simulations indicate that pressure can be used as a means to enhance the magnetic response of NiI$_{2}$. Even though studies of the induced electric polarization are beyond the scope of our work, we note that the anticipated decrease of the in-plane spiral pitch (as obtained in Appendix \ref{AppC}) is expected to enhance the spin-induced polarization (that should be proportional to the relative angle between neighboring spins as long as the spin-polarization tensor is not significantly affected by pressure). As such, pressure could also lead to an enhancement of the multiferroic properties of NiI$_{2}$. 


\section{Summary}

In this work, we have used first-principles calculations to explore the role of hydrostatic pressure in the structural, electronic, and magnetic response of bulk NiI$_{2}$. DFT-derived structural optimizations show good agreement with XRD data with the experimentally reported insulator-to-metal transition at $\sim$ 19 GPa being correctly reproduced by first-principles simulations. 
Using the four-state method, we have derived the intralayer and interlayer exchange parameters of NiI$_{2}$ (up to third nearest neighbors), finding a helimagnetic ground state with in-plane moments,  supported an easy-plane single-ion anisotropy and by the large magnetic frustration between the two dominant in-plane exchange terms ($J^{\parallel 1}$ and $J^{\parallel 3}$) of different sign (ferro- and antiferromagnetic, respectively). The interlayer exchanges are found to be antiferromagnetic with $J^{\perp 2}$ being dominant. As pressure is increased, $J^{\parallel 3}$ and $J^{\perp 2}$ become the overall dominant interactions with magnitudes that grow monotonically with pressure. This leads to our observation of the calculated bulk N\'{e}el temperatures increasing monotonically with pressure up to 10 GPa.  They saturate for higher pressures due to the competition between in-plane and out-of-plane couplings. Our results indicate that hydrostatic pressure is a promising way to enhance the magnetic response of NiI$_{2}$ and it could likely also be exploited to stabilize its multiferroic state at higher temperatures.   

\section{Acknowledgments}

We thank S. Picozzi for useful discussions during the early stages of this work. JK and AB acknowledge support from NSF Grant No. DMR 2206987 and the ASU Research Computing Center for high-performance computing resources. DA, BD and MJV acknowledge the SWIPE project funded by FNRS Belgium grant PINT-MULTI R.8013.20. MJV acknowledges ARC project DREAMS (G.A. 21/25-11) funded by Federation Wallonie Bruxelles and ULiege. PB acknowledges financial support from the Italian MIUR through Project No.PRIN 2017Z8TS5B. C.A.O., L.G.P.M., Q.S., and R.C. acknowledge support from the US Department of Energy, BES under Award No. DE-SC0019126 (materials synthesis and characterization and X-ray diffraction measurements).

\bibliography{references.bib}

\clearpage
\newpage
\onecolumngrid 

\appendix

\section{XRD Data and Extended Structural Parameter Calculations} \label{AppA}

\begin{figure}[H]
\includegraphics[width=\columnwidth]{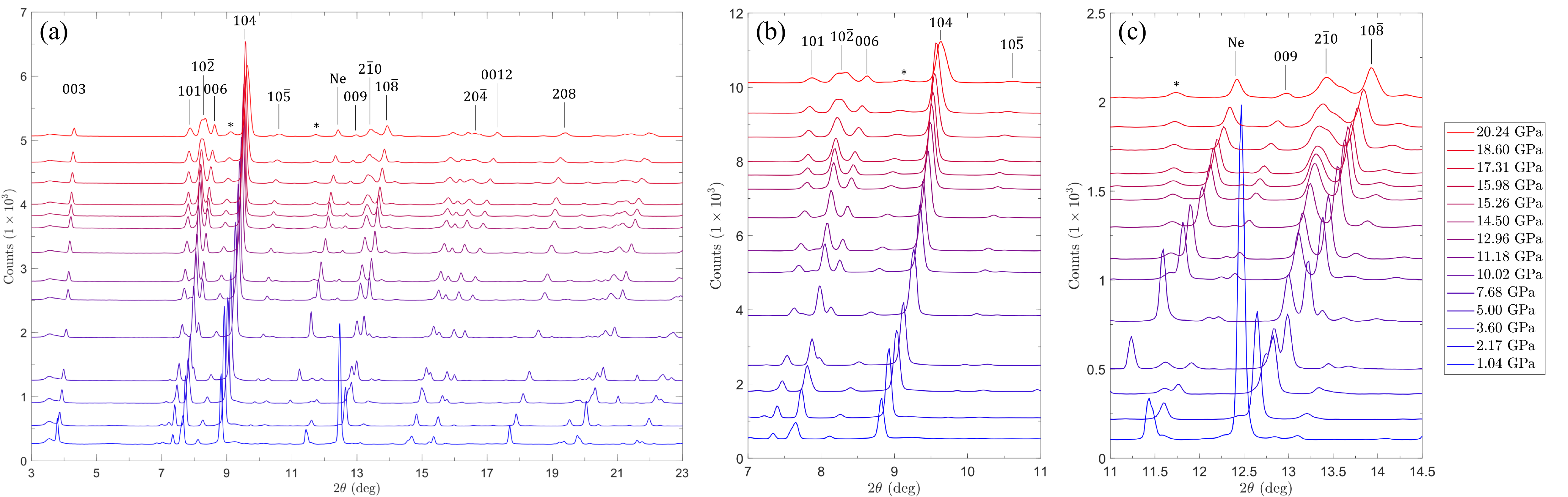}
\caption{(a) Bulk XRD spectrum at 200 K for increasing pressure (from $\sim$1 GPa in blue to $\sim$20 GPa in red). (b) and (c) show zoomed-in sections of the spectrum to highlight the splitting in the $10\overline{2}$ and $2\overline{1}0$ peaks, respectively, above $\sim$15 GPa.}
\label{fig:A1}
\end{figure}

High-pressure XRD measurements were performed at Sector 16-ID-B at the Advanced Photon Source, Argonne National Laboratory using an incident energy of $E=29.2$ keV. High-quality powder samples of NiI$_{2}$ grown by chemical vapor transport \cite{Song_Nat_2022} were loaded into a custom double-sided membrane-driven diamond anvil cell using Neon as a pressure transmitting medium with pressure monitored in-situ by ruby fluorescence. The results of the XRD measurements done at 200 K over a pressure range of $\sim$1 to $\sim$20 GPa are shown in Fig. \ref{fig:A1}. In-plane lattice parameter values can be extracted from either the $10\overline{2}$ or $2\overline{1}0$ peaks at each pressure (both of which undergo splitting above $\sim$15 GPa, indicative of the rhombohedral to monoclinic phase transition), while out-of-plane lattice parameters can be obtained from the $0012$ peaks. In-plane and out-of-plane lattice parameter values from the XRD measurements are given in Tables \ref{table:2} and \ref{table:3}, respectively.

In order to obtain the best agreement with the XRD lattice parameters, three Hubbard $U$ values (0, 3.24, and 5.8 eV), along with two exchange-correlation functionals were tested in the DFT calculations. The chosen test functionals were (i) a PBE exchange-correlation functional \cite{Perdew_PRL_1996} with added DFT-D3 vdW correction \cite{Grimme_JCP_2010} and (ii) a semilocal optB86b exchange-correlation functional \cite{Klimes_PRB_2011} augmented with a nonlocal PBE correlation functional. Tables \ref{table:2} and \ref{table:3} contain the 0 to 20 GPa results of the optimized in-plane and out-of-plane lattice parameters, respectively, for the bulk NiI$_{2}$ system in an AFM spin configuration. As can be seen in Table \ref{table:2}, for a particular pressure and $U$ value, the in-plane lattice parameters are not significantly affected by the choice of DFT functional. On the other hand, increasing $U$ gives rise to an increase in the lattice parameters. From Table \ref{table:3}, it is clear that the out-of-plane lattice parameter is, as expected, more sensitive to the choice of functional. Based on structural data alone, one would conclude that the PBE+D3 functional with a $U$ of 5.8 eV gives rise to the best agreement with experimental data. However, at this $U$ value the bulk system would still be insulating at 20 GPa, inconsistent with the known bulk insulator-to-metal transition at 19 GPa \cite{Pasternak_AIPCP_1994}. Thus, we proceed to use in the main text the structural data obtained for relaxations with PBE+D3 and $U=3.24$ eV, which still provide relatively good agreement with experimental lattice parameters, and also reproduce the experimentally observed insulator-to-metal transition with pressure. 

\renewcommand{\arraystretch}{1.3}
\begin{table}[H]
\centering
\begin{tabular}{lccccccc}
\hline
\hline
& \multicolumn{3}{c}{DFT: optB86b} & \multicolumn{3}{c}{DFT: PBE+D3} & \multirow{2}{*}{XRD} \\
$P$ (GPa) & $U=0.00$ & $U=3.24$ & $U=5.80$ & $U=0.00$ & $U=3.24$ & $U=5.80$ & \\
\hline
0 & 3.91 & 3.93 & 3.95 & 3.93 & 3.95 & 3.97 & -  \\
5 & 3.79 & 3.81 & 3.82 & 3.81 & 3.82 & 3.84 & 3.80 \\
10 & 3.70 & 3.73 & 3.73 & 3.67 & 3.73 & 3.74 & 3.72 \\
15 & 3.62 & 3.66 & 3.66 & 3.61 & 3.67 & 3.67 & 3.66 \\
20 & 3.56 & 3.61 & 3.61 & 3.56 & 3.61 & 3.61 & 3.63/3.59 \\
\hline
\hline
\end{tabular}
\caption{In-plane lattice parameters $a=b$ (in units of \AA) for bulk NiI$_{2}$ from DFT optimization calculations (using an AFM spin configuration) and XRD measurements at pressures ($P$) of 0 to 20 GPa. DFT results are shown for two functionals, optB86b and PBE+D3, for which three Hubbard $U$ values were tested: 0, 3.24, and 5.8 eV (with corresponding Hund's $J_{\mathrm{H}}$ values of 0, 0.68, and 0.8 eV, respectively). XRD results are given for measurements taken at 200 K. As mentioned in the main text, above $\sim$15 GPa there is a phase transition to monoclinic symmetry so both values for the in-plane lattice constant are provided for the 20 GPa case.}
\label{table:2}
\end{table}

\renewcommand{\arraystretch}{1.3}
\begin{table}[H]
\centering
\begin{tabular}{lccccccc}
\hline
\hline
& \multicolumn{3}{c}{DFT: optB86b} & \multicolumn{3}{c}{DFT: PBE+D3} & \multirow{2}{*}{XRD} \\
$P$ (GPa) & $U=0.00$ & $U=3.24$ & $U=5.80$ & $U=0.00$ & $U=3.24$ & $U=5.80$ & \\
\hline
0 & 18.83 & 19.44 & 19.69 & 19.07 & 19.72 & 19.99 & - \\
5 & 17.60 & 18.08 & 18.35 & 17.56 & 18.11 & 18.41 & 18.31 \\
10 & 17.07 & 17.50 & 17.77 & 16.94 & 17.45 & 17.76 & 17.69 \\
15 & 16.74 & 17.10 & 17.39 & 16.64 & 17.03 & 17.35 & 17.31 \\
20 & 16.52 & 16.78 & 17.10 & 16.39 & 16.66 & 17.04 & 16.92 \\
\hline
\hline
\end{tabular}
\caption{Out-of-plane lattice parameters $c$ (in units of \AA) for bulk NiI$_{2}$ from DFT optimization calculations (using an AFM spin configuration) and XRD measurements at pressures ($P$) of 0 to 20 GPa. DFT results are shown for two functionals, optB86b and PBE+D3, for which three Hubbard $U$ values were tested: 0, 3.24, and 5.8 eV (with corresponding Hund's $J_{\mathrm{H}}$ values of 0, 0.68, and 0.8 eV, respectively). XRD results are given for measurements taken at 200 K.}
\label{table:3}
\end{table}

\section{Comparison between Bulk and Bilayer and Extended Band Structure Calculations} \label{AppB}

\renewcommand{\arraystretch}{1.3}
\begin{table}[H]
\centering
\begin{tabular}{lcccccc}
\hline
\hline
 & \multicolumn{3}{c}{Bilayer} & \multicolumn{3}{c}{Bulk} \\
$P$ (GPa) & $\Delta E$ (meV) & $J^{\perp \mathrm{eff}}$ (meV) & $E_{\mathrm{gap}}^{\mathrm{AFM}}$ (eV) & $\Delta E$ (meV) & $J^{\perp \mathrm{eff}}$ (meV) & $E_{\mathrm{gap}}^{\mathrm{AFM}}$ (eV) \\
\hline
0 & 6.8 & 2.3 & 0.99 & 13.6 & 2.3 & 0.98 \\
5 & 23.7 & 7.9 & 0.75 & 47.3 & 7.9 & 0.70 \\
10 & 41.1 & 13.7 & 0.54 & 81.5 & 13.6 & 0.40 \\
15 & 59.2 & 19.7 & 0.33 & 105.4 & 17.6 & 0.14 \\
20 & 74.4 & 24.8 & 0.14 & 108.7 & 18.1 & 0.00 \\
\hline
\hline
\end{tabular}
\caption{Energy difference between FM and AFM states ($\Delta E$) along with the corresponding effective interlayer exchange ($J^{\perp \mathrm{eff}}$) for bilayer and bulk NiI$_{2}$ at pressures ($P$) of 0 to 20 GPa. AFM band gap energies ($E_{\mathrm{gap}}^{\mathrm{AFM}}$) for bilayer and bulk NiI$_{2}$ are also given.}
\label{table:4}
\end{table}

Table \ref{table:4} shows the energy difference (per formula unit) between the FM and AFM spin configurations for bilayer and bulk NiI$_{2}$: $\Delta E = E_{\mathrm{FM}} - E_{\mathrm{AFM}}$. A simple Heisenberg Hamiltonian $H = \left( J^{\perp \mathrm{eff}}/2 \right) \sum_{i,j} \mathbf{S}_{i} \cdot \mathbf{S}_{j}$ captures the effective interlayer exchange interaction, where $J^{\perp \mathrm{eff}}$ is the isotropic exchange constant between spins $\mathbf{S}_{i,j}$ on out-of-plane first-nearest neighbor Ni lattice sites $i,j$. Again, the factor of 1/2 in front of the summation is to account for double-counting. In the bilayer (bulk) case there are 3 (6) nearest out-of-plane neighbors, and so we obtain the following energy equations for the two spin states: $E_{\mathrm{FM}}^{\mathrm{bilayer}} = E_{0} + (3/2) J^{\perp \mathrm{eff}} S^{2}$ ($E_{\mathrm{FM}}^{\mathrm{bulk}} = E_{0} + 3 J^{\perp \mathrm{eff}} S^{2}$) and $E_{\mathrm{AFM}}^{\mathrm{bilayer}} = E_{0} - (3/2) J^{\perp \mathrm{eff}} S^{2}$ ($E_{\mathrm{AFM}}^{\mathrm{bulk}} = E_{0} - 3 J^{\perp \mathrm{eff}} S^{2}$), where $E_{0}$ is the total energy for the system omitting magnetic interactions and note that $S^{2}=|\mathbf{S}|^{2}$. Taking the difference yields $\Delta E^{\mathrm{bilayer}} = E_{\mathrm{FM}}^{\mathrm{bilayer}} - E_{\mathrm{AFM}}^{\mathrm{bilayer}} = 3 J^{\perp \mathrm{eff}} S^{2}$ and $\Delta E^{\mathrm{bulk}} = E_{\mathrm{FM}}^{\mathrm{bulk}} - E_{\mathrm{AFM}}^{\mathrm{bulk}} = 6 J^{\perp \mathrm{eff}} S^{2}$. Finally, solving for the interlayer exchange in each case, we have $J^{\perp \mathrm{eff}} = (E_{\mathrm{FM}}^{\mathrm{bilayer}} - E_{\mathrm{AFM}}^{\mathrm{bilayer}}) / 3S^{2}$ and $J^{\perp \mathrm{eff}} = (E_{\mathrm{FM}}^{\mathrm{bulk}} - E_{\mathrm{AFM}}^{\mathrm{bulk}}) / 6S^{2}$. With $S_{\mathrm{Ni}}=1$, we obtain the bilayer and bulk $J^{\perp \mathrm{eff}}$ values given in Table \ref{table:4}, where at each pressure $\Delta E > 0$ indicating that the ground state is AFM for both systems. Subsequently, $J^{\perp \mathrm{eff}} > 0$ indicating an AFM exchange. Further, note that the exchange is nearly identical in both systems up to 10 GPa, after which the bilayer exchange becomes larger than in the bulk. In both systems, however, we see a trend of monotonically increasing exchange with pressure, as one would expect considering the decreasing interlayer distance. 

\begin{figure}[H]
\includegraphics[width=\columnwidth]{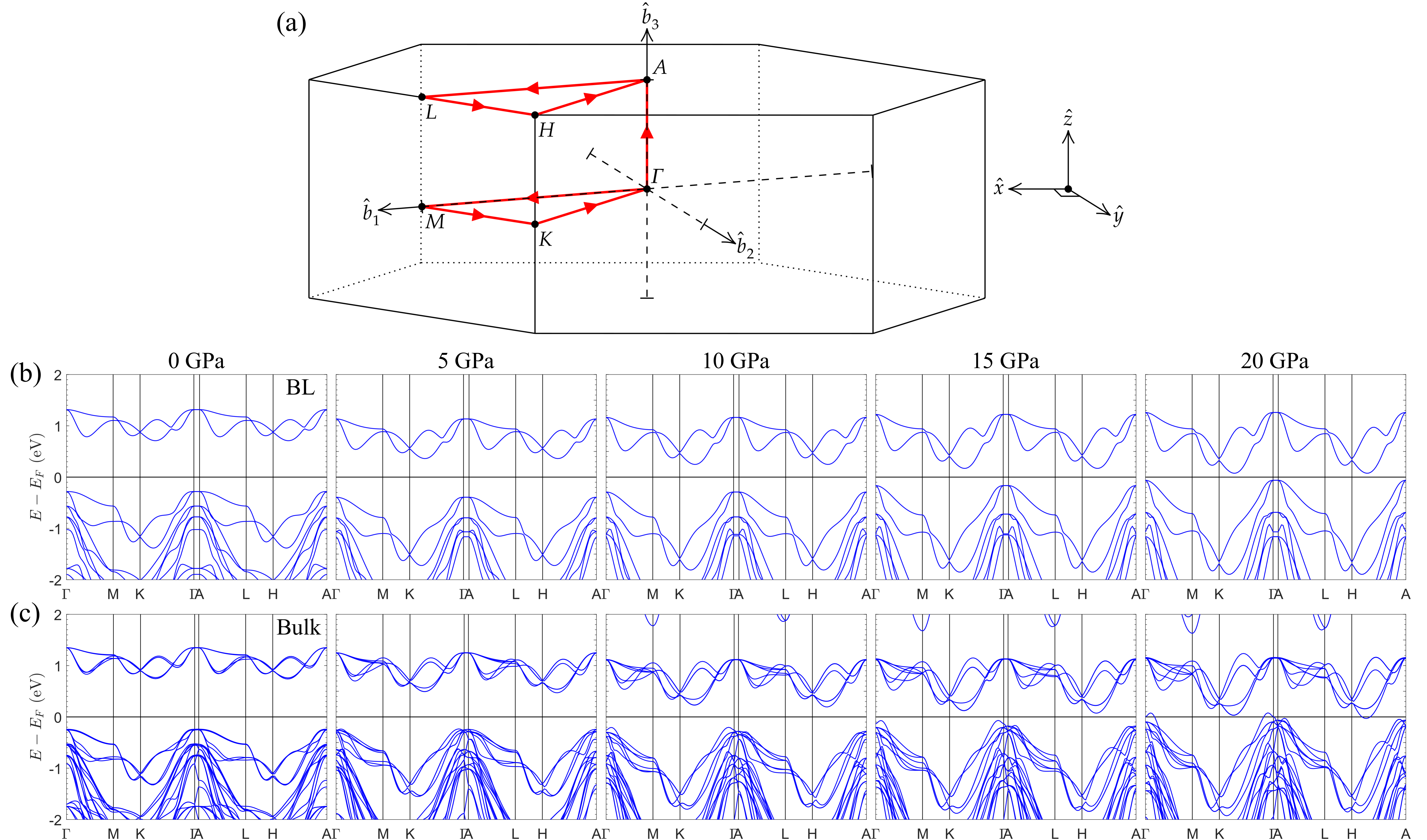}
\caption{(a) First Brillouin zone for NiI$_{2}$ where high-symmetry points are labeled and the chosen $k$-path for our electronic structure calculations is indicated by the red arrows. The high symmetry points correspond to: $\Gamma=(0,0,0)$, $\mathrm{M}=(1/2,0,0)$, $\mathrm{K}=(1/3,1/3,0)$, $\mathrm{A}=(0,0,1/2)$, $\mathrm{L}=(1/2,0,1/2)$, $\mathrm{H}=(1/3,1/3,1/2)$. (b) Electronic structure for bilayer NiI$_{2}$ in an AFM spin configuration from 0-20 GPa. (c) Electronic structure for bulk NiI$_{2}$ in an AFM spin configuration from 0-20 GPa.}
\label{fig:B1}
\end{figure}

Table \ref{table:4} also shows the evolution with pressure of the band gap energies for bulk and bilayer NiI$_{2}$ for an AFM state (consisting of ferromagnetic planes coupled antiferromagnetically out-of-plane). The corresponding electronic band structures are shown in Fig. \ref{fig:B1} (the first BZ is illustrated in Fig. \ref{fig:B1}a where the chosen k-path is indicated). A decrease in the band gap with increasing pressure can be observed along with an increase in the bandwidths. This is expected since the out-of-plane lattice parameter, and thus the interlayer distance, decreases as pressure increases. In the bilayer case, the system remains insulating up to 20 GPa, while the bulk transitions to a metal between 15 and 20 GPa, consistent with the known insulator-to-metal transition of 19 GPa. In the bilayer band structures, flat bands between $\Gamma$ and A at each pressure can be observed as well as nearly identical $k_{z}=0$ and $k_{z}=0.5$ band structures, implying that the bilayer is indeed isolated. In contrast, significantly dispersive bands can be observed in the bulk between $\Gamma$ and A (with the dispersion increasing with pressure), and also there is clear variation between the $k_{z}=0$ and $k_{z}=0.5$ band structures.

\begin{figure}[H]
\includegraphics[width=\columnwidth]{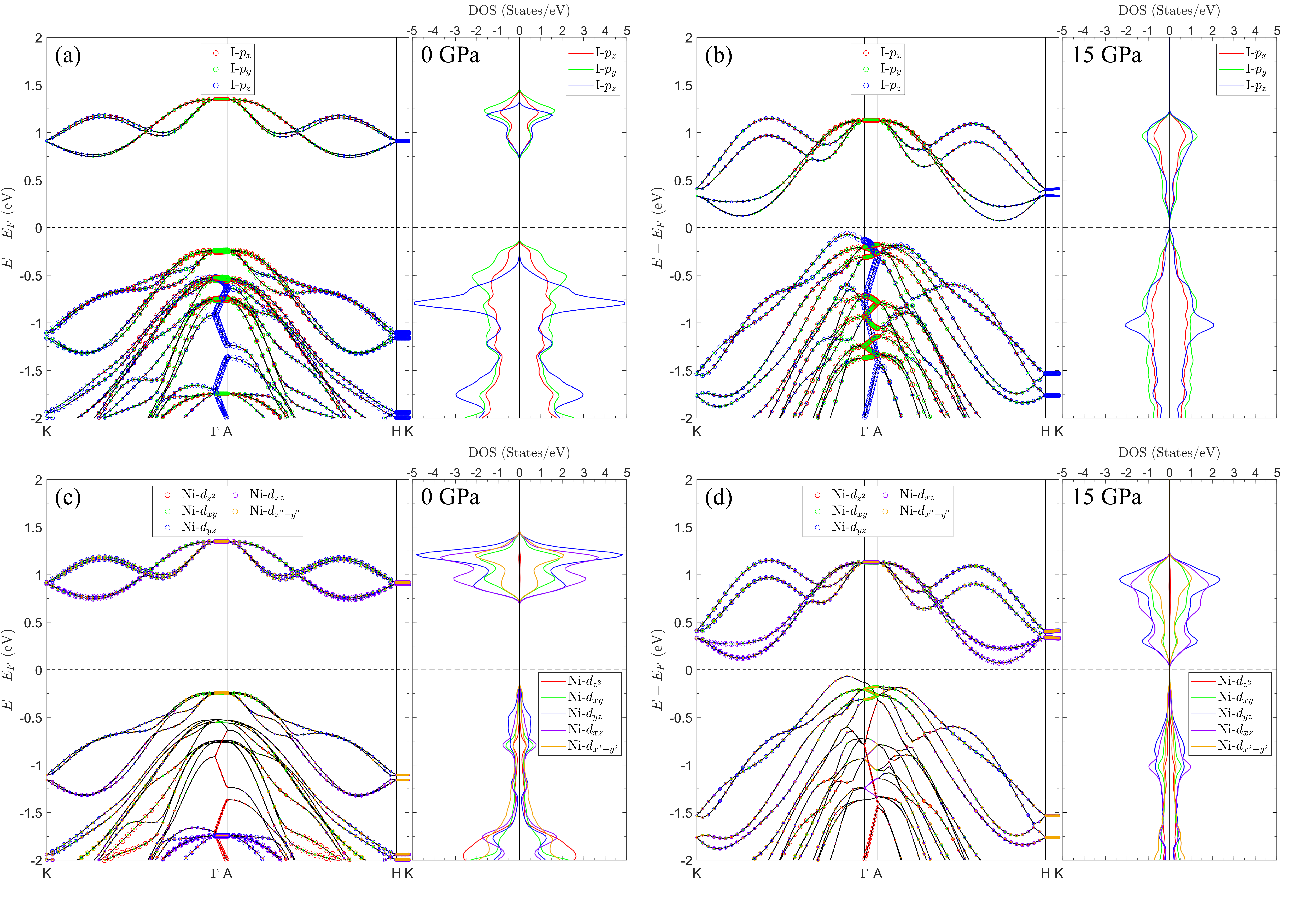}
\caption{Band character plots and orbital-resolved density of states for AFM bulk NiI$_{2}$. (a) I-$p$ states at 0 GPa. (b) I-$p$ states at 15 GPa. (c) Ni-$d$ states at 0 GPa. (d) Ni-$d$ states at 15 GPa. Note that here we use a different k-path than in the previous electronic structures in Fig. \ref{fig:B1} to better show the band dispersion from $\Gamma$ to A.}
\label{fig:B2}
\end{figure}

To determine the orbital character of these dispersive bands, we look at the orbital-resolved density of states (DOS) and also the band character plots for bulk NiI$_{2}$ as shown in Fig. \ref{fig:B2}. The I-$p$ states at 0 and 15 GPa are shown in Figs. \ref{fig:B2}a and \ref{fig:B2}b, respectively, while the Ni-$d$ states are shown in Figs. \ref{fig:B2}c and \ref{fig:B2}d, respectively. The I-$p$ states lie directly below the Fermi level and have partial hybridization with the Ni-$d$ states which are lower in energy. Clearly, the highly dispersive bands between $\Gamma$ and A come mostly from the I-$p_{z}$ states, but they have Ni-$d_{z^{2}}$ admixture as well.


\section{Further Details on the Exchange Interactions} \label{AppC}

The exchange interaction tensor can be decomposed into contributions from isotropic as well as anisotropic exchanges. The isotropic component is given by $J_{ij}^{\parallel} = \frac{1}{3} \mathrm{Tr} \left[ \mathbf{J}_{ij}^{\parallel} \right]$, where Tr is the trace operator over spatial directions. The anisotropic contributions can be further decomposed into antisymmetric and symmetric components, ($\mathbf{J}_{ij}^{\mathrm{A} \parallel}$ and $\mathbf{J}_{ij}^{\mathrm{S} \parallel}$, respectively), 
the effects of which will tend to favor either an alignment of the spins in a given direction, as in the symmetric case, or a canting of spin pairs, as in the antisymmetric case. The antisymmetric exchange corresponds to the Dzyaloshinskii–Moriya (DM) interaction, which is zero for an inversion center at the point bisecting the Ni-Ni bond, as in the present case \cite{Amoroso_NC_2020,Moriya_PR_1960,Simon_JPCM_2014,Vida_PRB_2016}.
The symmetric exchange (two-site anisotropy) on the other hand, has the form $\mathbf{J}_{ij}^{\mathrm{S} \parallel} = \frac{1}{2} \left( \mathbf{J}_{ij}^{\parallel} + \left[ \mathbf{J}_{ij}^{\parallel} \right]^{\mathrm{T}} \right) - J_{ij}^{\parallel} \mathbf{I}$ (here $\mathbf{I}$ is the unit matrix and the superscript $\mathrm{T}$ denotes the transpose) and can play an important role in the spontaneous stabilization of noncollinear magnetic states \cite{Amoroso_NC_2020}. For a Ni-Ni pair whose bond is along the Cartesian $\hat{x}$ axis, the exchange tensor has the form \cite{Amoroso_NC_2020}
\begin{equation*}
	\mathbf{J}^{\parallel} = 
		\left(
		\begin{matrix}
		J^{\parallel}_{xx} & 0 & 0 \\
		0 & J^{\parallel}_{yy} & J^{\parallel}_{yz} \\
		0 & J^{\parallel}_{zy} & J^{\parallel}_{zz}
		\end{matrix}
		\right) \, .
\end{equation*}
\noindent Note that, by symmetry, $J^{\parallel}_{yz} = J^{\parallel}_{zy}$ and the other off-diagonal terms are zero in the absence of DMI. Lastly, the SIA tensor in Eq. \ref{eq:1} can be simplified due to the local symmetry of Ni ions which has the Cartesian $\hat{z}$-direction parallel with the crystallographic $c$ axis, hence $H^{\mathrm{SIA}}= \sum_{{i}} A_{i} \left( S_{i,z} \right)^{2}$, where $A_{i}$ is now a scalar \cite{Amoroso_NC_2020}.

\begin{figure}[H]
\includegraphics[width=\columnwidth]{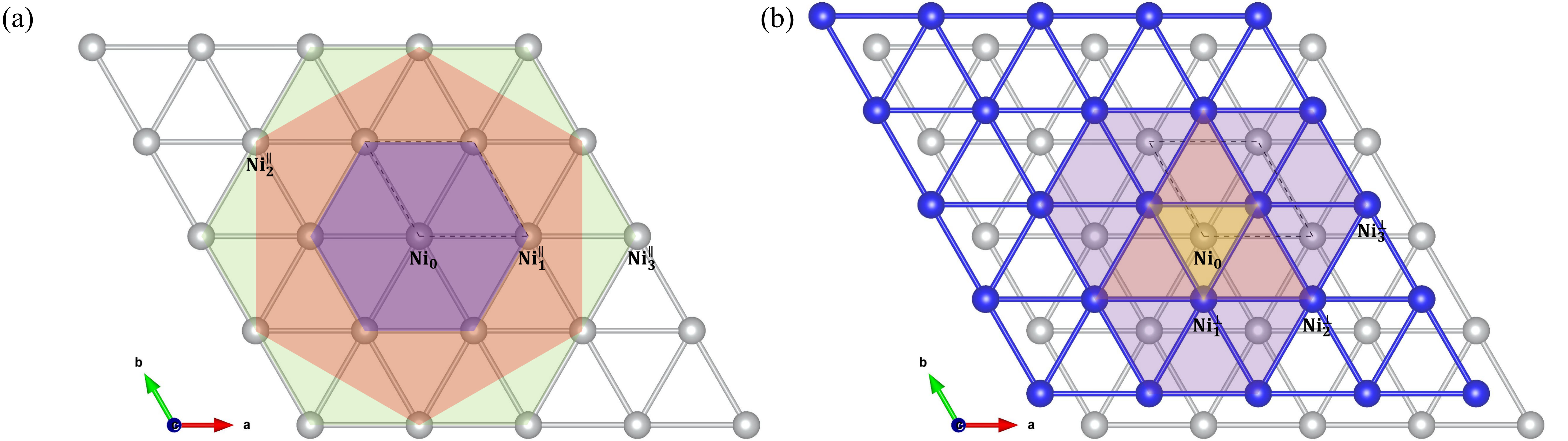}
\caption{(a) Triangular lattice of Ni atoms (gray spheres) in monolayer NiI$_{2}$ illustrating intralayer nearest neighbors. Ni$_{0}$ is the central atom and the Ni atoms at the corners of the blue, red, and green hexagons are the first, second, and third in-plane nearest neighbors of Ni$_{0}$, respectively. (b) Triangular lattice of Ni atoms (gray spheres in the bottom layer and blue spheres in the top layer) in bilayer NiI$_{2}$ illustrating the interlayer nearest neighbors. Ni$_{0}$ is the central atom and the Ni atoms at the corners of the 
yellow and orange triangles are the first and second out-of-plane nearest neighbors of Ni$_{0}$, respectively, while the atoms at the corners of the purple polygon are the third out-of-plane nearest neighbors of Ni$_{0}$. In (a) and (b) the unit cells are indicated by black dashed lines.}
\label{fig:C1}
\end{figure}

In Fig. \ref{fig:C1} we show the intra- and interlayer nearest neighbors for mono- and bilayer NiI$_{2}$ lattices. In each system, the nearest neighbor Ni atom pairs chosen for the four-state calculations are indicated. In the monolayer, the $\mathrm{Ni}_{0}-\mathrm{Ni}_{1}^{\parallel}$ pair is chosen for the first nearest neighbor interaction ($\mathbf{J}^{\parallel 1}$), the $\mathrm{Ni}_{0}-\mathrm{Ni}_{2}^{\parallel}$ pair is chosen for the second nearest neighbor interaction ($J^{\parallel 2}$), and the $\mathrm{Ni}_{0}-\mathrm{Ni}_{3}^{\parallel}$ pair is chosen for the third nearest neighbor interaction ($J^{\parallel 3}$). In the bilayer, the $\mathrm{Ni}_{0}-\mathrm{Ni}_{1}^{\perp}$ pair is chosen for the first nearest neighbor interaction ($J^{\perp 1}$), the $\mathrm{Ni}_{0}-\mathrm{Ni}_{2}^{\perp}$ pair is chosen for the second nearest neighbor interaction ($J^{\perp 2}$), and the $\mathrm{Ni}_{0}-\mathrm{Ni}_{3}^{\perp}$ pair is chosen for the third nearest neighbor interaction ($J^{\perp 3}$).

Table \ref{table:5} contains the ratio of $J^{\parallel 1}$ and $J^{\parallel 3}$ from the four-state calculations, which is used to obtain the magnetic propagation vector $q = |\mathbf{q}| = 2\arccos{\left[ \left( 1 + \sqrt{1 - 2 (J^{\parallel 1}/J^{\parallel 3}) } \right) / 4 \right]}$ \cite{Hayami_PRB_2016,Batista_RPP_2016}. This can then be used to obtain the size of the magnetic unit cell $L_{\mathrm{m.u.c.}} \sim 4 \pi /q$. A decrease in the $J^{\parallel 1}/J^{\parallel 3}$ ratio with pressure leads to an increase in $q$, which results in a shorter pitch of the in-plane spin-spiral and consequently a smaller magnetic unit cell length $L_{\mathrm{m.u.c.}}$. 

\renewcommand{\arraystretch}{1.3}
\begin{table}[H]
\centering
\begin{tabular}{lccc}
\hline
\hline
$P$ (GPa) & $J^{\parallel 1}/J^{\parallel 3}$ & $q$ & $L_{\mathrm{m.u.c.}}$ \\
\hline
0 & -1.23 & 0.12 & 8.12 \\
5 & -0.85 & 0.14 & 7.40 \\
10 & -0.62 & 0.14 & 7.01 \\
15 & -0.46 & 0.15 & 6.75 \\
20 & -0.36 & 0.15 & 6.58 \\
\hline
\hline
\end{tabular}
\caption{Ratio of leading intralayer exchanges $J^{\parallel 1}/J^{\parallel 3}$ in bulk NiI$_{2}$ at pressures $P$ up to 20 GPa with the corresponding magnetic propagation vector magnitude $q$ and magnetic unit cell length $L_{\mathrm{m.u.c.}}$.}
\label{table:5}
\end{table}




\section{Extended MC calculations} \label{AppD}

\begin{figure}[H]
\includegraphics[width=\columnwidth]{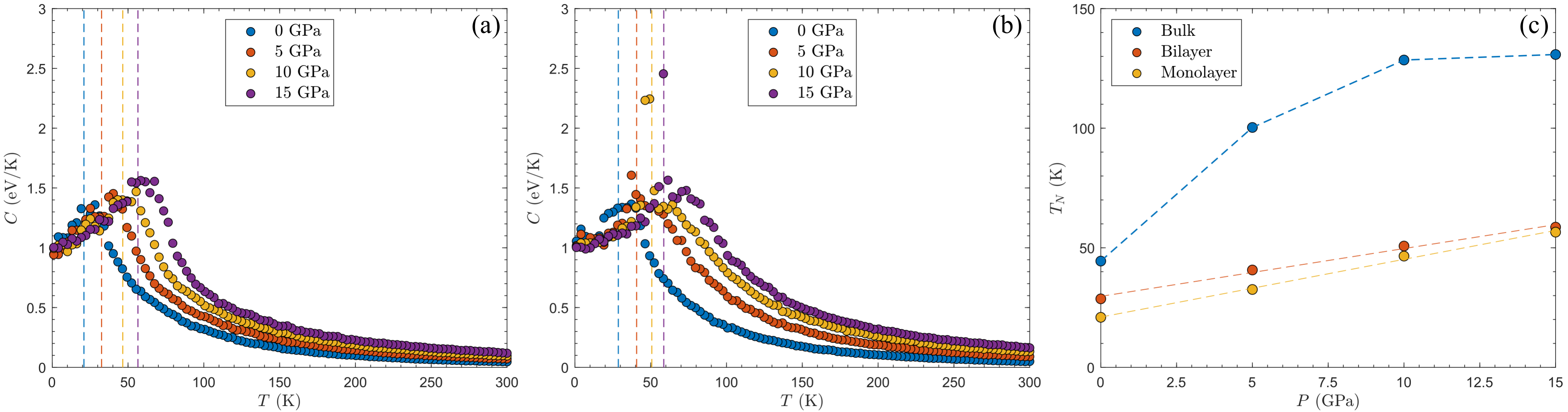}
\caption{(a) Monolayer and (b) bilayer NiI$_{2}$ specific heat $C$ as a function of temperature $T$ for pressures of 0, 5, 10, and 15 GPa (blue, orange, yellow, and purple circles, respectively) obtained from Monte Carlo simulations. The dashed lines indicate the N\'{e}el temperature at each pressure (obtained via curve fitting with a general Lorentzian function). (c) N\'{e}el temperature values for bulk, bilayer, and monolayer NiI$_{2}$ from MC calculations (blue, orange, and yellow circles, respectively) as a function of pressure $P$. The dashed lines represent linear best-fits to the MC data (except in the bulk case where the dashed line simply connects the data points to serve as a guide for the eye).}
\label{fig:D1}
\end{figure}

In Fig. \ref{fig:D1} we show the specific heat calculated from MC as a function of temperature for the mono- and bilayer NiI$_{2}$ systems as a function of pressure (up to 15 GPa). The N\'{e}el temperature at each pressure was obtained by fitting the specific heat vs. temperature plot using a general Lorentzian curve of the form $C(T) = f(x) = C_{0} + \frac{2A}{\pi} \left( \frac{w}{w^{2} + 4(x-x_{c})^{2}} \right)$, where $A$ is the area under the curve, $w$ is the width of the peak, $x_{c}$ is the peak maximum, $C_{0}$ is the offset, and $[A,w,x_{c},C_{0}] \in \mathbb{R}$. Here, the N\'{e}el temperature $T_{\mathrm{N}}$ is equal to $x_{c}$. The particular N\'{e}el temperature values for monolayer, bilayer, and bulk NiI$_{2}$ from 0-15 GPa are given in Table \ref{table:6} and also plotted vs. pressure in Fig. \ref{fig:D1}c. In mono- and bilayers, $T_{\mathrm{N}}$ increases monotonically with pressure but a monotonic increase in $T_{\mathrm{N}}$ with layer number can also be observed, which is consistent with the literature \cite{Song_Nat_2022}. The mono- and bilayer ambient pressure $T_{\mathrm{N}}$ values themselves are consistent with the literature with a N\'{e}el temperature of 21 K and 30 K reported for monolayer and bilayer NiI$_{2}$, respectively \cite{Song_Nat_2022}. While both mono- and bilayer N\'{e}el temperatures increase linearly with pressure up to 15 GPa (with slopes of 2.4 and 2.0 K/GPa, respectively) the bulk values appear to saturate around $\sim$130 K above 10 GPa, 
the cause of which is examined in Appendix \ref{AppE}.

\renewcommand{\arraystretch}{1.3}
\begin{table}[H]
\centering
\begin{tabular}{lccc}
\hline
\hline
$P$ (GPa) & $T_{\mathrm{N}}^{\mathrm{ML}}$ (K) & $T_{\mathrm{N}}^{\mathrm{BL}}$ (K) & $T_{\mathrm{N}}^{\mathrm{Bulk}}$ (K) \\
\hline
0 & 20.9 & 28.7 & 44.4 \\
5 & 32.6 & 40.7 & 100.3 \\
10 & 46.5 & 50.7 & 128.5 \\
15 & 56.5 & 58.6 & 130.8 \\
\hline
\hline
\end{tabular}
\caption{N\'{e}el temperatures ($T_{\mathrm{N}}$) for monolayer (ML), bilayer (BL), and bulk NiI$_{2}$ systems from MC calculations at pressures ($P$) of 0 to 15 GPa.}
\label{table:6}
\end{table}


\section{Mean-Field Approach to NiI\texorpdfstring{$_{2}$}{2} Critical Temperature Under Pressure} \label{AppE}

To further investigate the saturation of $T_{\mathrm{N}}$ above 10 GPa, we estimate the N\'{e}el temperature of NiI$_{2}$ at different pressures from a mean-field standpoint. First, we derive the mean-field estimate for $T_{\mathrm{N}}$ assuming a simple AFM configuration of FM layers coupled antiferromagnetically. For the sake of simplicity, we consider a minimal Heisenberg model with only isotropic $J^{\parallel 1}$, $J^{\parallel 3}$, and an effective interlayer coupling $J^{\perp \mathrm{eff}}$. In a mean-field approach, one has to derive an effective Weiss field $B_{i}$, generated from neighboring spins, acting on a target spin $S_{i}$. For classical spins, the expectation value of the local magnetization is then derived from
\begin{equation} \label{eq:E1}
    M_{i} = \langle S_{i} \rangle = \coth{\left( \beta B_{i} \right)} - \frac{1}{\beta B_{i}} \approx \frac{\beta B_{i}}{3} \, ,
\end{equation}
\noindent where $\beta = 1/k_{B}T$ with $T$ being the temperature and $k_{B}$ the Boltzmann constant. The last approximated equality is valid in the limit of vanishing Weiss field, i.e. close to the transition of the sought ordered phase. As the Weiss field depends on the magnetization, Eq. \ref{eq:E1} represents a self-consistent equation for the critical temperature.

In the simple AFM configuration, we can distinguish two magnetic sub-lattices $M_{1}$ and $M_{2}$, each defined in alternating layers (e.g., $M_{1}$ in odd layers and $M_{2}$ in even ones). Then, the AFM order parameter is defined simply as $L = M_{1} - M_{2}$, while the FM order parameter is $M_{1} + M_{2}$. Taking into account the rhombohedral structure of NiI$_{2}$, the Weiss fields are thus
\begin{equation} \label{eq:E2}
\begin{split}
    B_{1} & = - J_{\parallel} M_{1} - J_{\perp} M_{2} \, , \\
    B_{2} & = - J_{\parallel} M_{2} - J_{\perp} M_{1} \, ,
\end{split}
\end{equation}
\noindent where $J_{\parallel} = 6 J^{\parallel 1} + 6 J^{\parallel 3}$ and $J_{\perp} = 6 J^{\perp \mathrm{eff}}$. Combining Eqs. \ref{eq:E1} and \ref{eq:E2} and using the definition of order parameters, the following expressions are obtained:
\begin{subequations}
\begin{align}
M &= - \frac{\beta}{3} \left( J_{\parallel} + J_{\perp} \right) M \, , \label{eq:E3a} \\
L &= - \frac{\beta}{3} \left( J_{\parallel} - J_{\perp} \right) L \, . \label{eq:E3b}
\end{align}
\label{eq:E3}
\end{subequations}
\noindent Using the last equation, the mean-field estimate for the AFM critical temperature is thus $3 k_{B} T_{c} = J_{\perp} - J_{\parallel}$ (note that it is positive, as negative/positive exchange constants denote FM/AFM interactions).

\begin{figure}[H]
\includegraphics[width=\columnwidth]{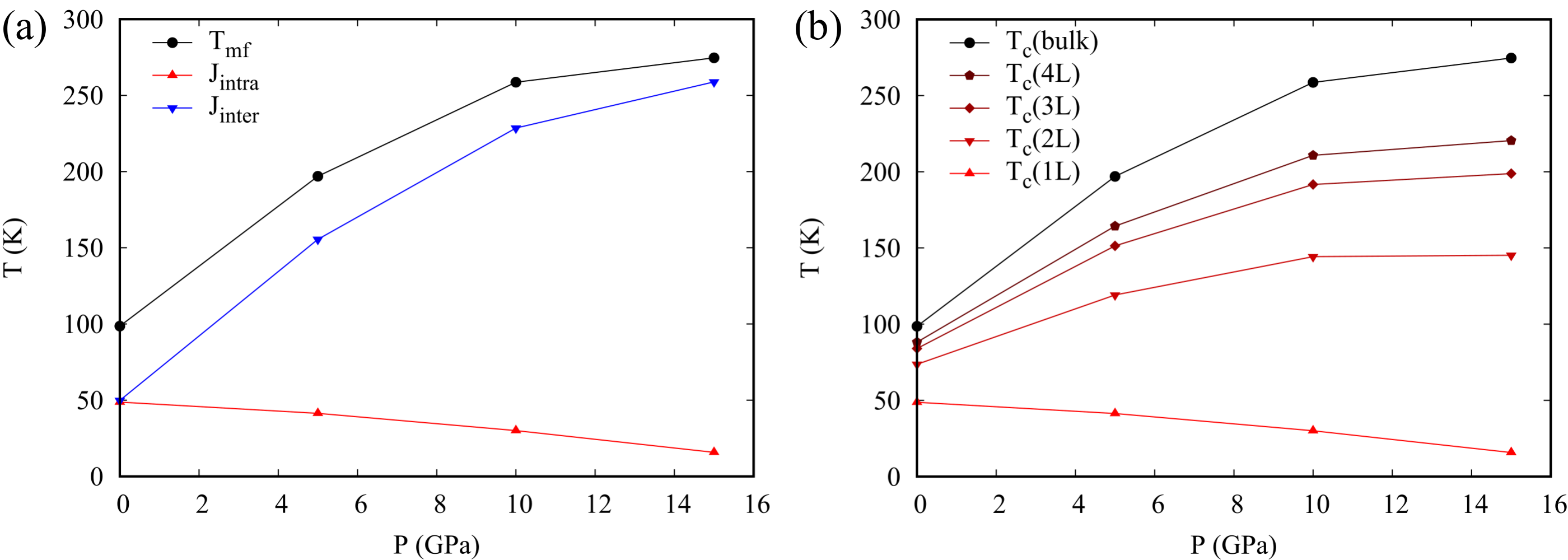}
\caption{(a) Evolution of the mean-field (mf) estimate of $T_{c}$ under pressure using the exchange parameters $J^{\perp \mathrm{eff}}$, $J^{\parallel 1}$, and $J^{\parallel 3}$ taken from Table \ref{table:1} in the main text. The labels $\mathrm{J}_{\mathrm{intra}}$ and $\mathrm{J}_{\mathrm{inter}}$ denote the intralayer $-J_{\parallel}(q)$ and interlayer $J_{\perp}(q)$ contributions to $\mathrm{T}_{\mathrm{mf}}$, respectively. (b) Evolution of mean-field estimate of $T_{c}$ under pressure for different number of layers, from the monolayer (intralayer contribution only) to 4 layers, with same parameters used in (a). The prediction for the bulk $T_{c}$ is also shown as a reference.}
\label{fig:E1}
\end{figure}

The critical temperature for the helimagnetic configuration can be similarly deduced, imposing that the spins surrounding the targeted one (for which we want to identify the Weiss field) are arranged according to the magnetic propagation vector. We assume the in-plane propagation vector with the value of $q$ that minimizes the classical energy, i.e. $q= 2\arccos{\left[ \left( 1 + \sqrt{1 - 2 (J^{\parallel 1}/J^{\parallel 3}) } \right) / 4 \right]}$. Now the Weiss fields can still be formally written as in Eq. \ref{eq:E2}, simply replacing
\begin{equation} \label{eq:E4}
\begin{split}
    J_{\parallel} &\mapsto J_{\parallel}(q) = 4 J^{\parallel 1} \cos{\left( \frac{q}{2} \right)} + \left( 2 J^{\parallel 1} + 4 J^{\parallel 3} \right) \cos{\left( q \right)} + 2 J^{\parallel 3} \cos{\left( 2q \right)} \, , \\
    J_{\perp} &\mapsto J_{\perp}(q) = 2 J^{\perp \mathrm{eff}} \Bigl( 1 + \cos{\left( q \right)} + \cos{\left( 2q \right)} \Bigr)  \, .
\end{split}
\end{equation}
\noindent It follows that the mean-field estimate for the helimagnetic configuration with AFM modulation along the $c$ axis is formally given by Eq. \ref{eq:E3b} upon replacements of Eqs. \ref{eq:E4}, that is
\begin{equation} \label{eq:E5}
    3 k_{B} T_{c} = J_{\perp}(q) - J_{\parallel}(q) \, .
\end{equation}
\noindent Despite the severe approximations introduced by i) the mean-field Weiss approach and ii) simplifying the model to only three interactions (including an effective interlayer one) and also neglecting anisotropies, the mean-field prediction works surprisingly well, and most importantly it reproduces the saturation behaviour under pressure, as shown in Fig. \ref{fig:E1}a. This stems from the interplay of interlayer and intralayer terms, with the strongest effect seeming to come from $J_{\perp}(q)$ (which is a function not only of $J^{\perp \mathrm{eff}}$, but $J^{\parallel 1}$ and $J^{\parallel 3}$ as well), as the optimal $q$ depends explicitly on the Heisenberg exchange interactions also. Therefore, the saturation seems to arise from a competition between all interactions, and not merely between the dominant $J^{\perp \mathrm{eff}}$ and $J^{\parallel 3}$.

Although the mean-field approach largely overestimates $T_{c}$, a well-known consequence of neglecting fluctuations (in this case it may also come from the simplified model, where we neglected $J^{\parallel 2}$ as well as the real topology of interlayer couplings $J^{\perp 1}$, $J^{\perp 2}$, and $J^{\perp 3}$), the order of magnitude of the pressure-induced enhancement of $T_{c}$ is consistent with the MC simulations shown in Fig. \ref{fig:5} (i.e. roughly a three-fold increase at the largest considered pressure). On the other hand, the intralayer contribution to $T_{c}$ is found to decrease in this mean-field approach, which is inconsistent with the MC results (the inclusion of the second-nearest neighbor interaction partly corrects the trend). 

Lastly, one can in principle adopt the same mean-field scheme to address the dependence on the number of layers. The mean-field formulas are consistent with previous MC calculations in showing that
\begin{equation} \label{eq:E6}
    \frac{T_{c}^{\mathrm{bulk}}}{T_{c}^{\mathrm{1L}}} = 1 + \frac{J_{\perp}(q)}{J_{\parallel}(q)} = 1 + \frac{J^{\perp \mathrm{eff}}}{J^{\parallel 1}} \rho \left( q, \frac{J^{\parallel 3}}{J^{\parallel 1}} \right) \, .
\end{equation}

\noindent The mean-field temperatures for $L=2,3,4$ were also calculated, but in these cases the expression for a general $L$ can not be extrapolated as one has to find an iterative solution for the self-consistent Weiss equations. Trends in $T_{c}$ for $L=2,3,4$ are shown in Fig. \ref{fig:E1}b and are roughly consistent with expectations but with same issues discussed previously.

\end{document}